\documentstyle[preprint,aps]{revtex}
\tighten
\begin{document}

\maketitle
\noindent 
{\bf Comment on "Observation of Superluminal Behaviors in  
Wave Propagation"}
\\
\\
\\
\indent\indent\indent\indent\indent{\bf Murray Peshkin} \\
\\
\indent\indent\indent\indent\indent {\it Argonne National Laboratory} \\ 
\indent\indent\indent\indent\indent {\it Physics Division-203} \\
\indent\indent\indent\indent\indent {\it Argonne, IL 60439-4843} \\
\indent\indent\indent\indent\indent {\it peshkin@anl.gov} \\
\\
\\
\\
Mugnai {\it et\,al.} [1] have reported an experiment in which microwave
packets appear to travel in air with a speed substantially greater than $c$, the
normal speed of light in vacuum.  The packets consist of square pulses, many
wavelengths in extent, which are chopped segments of the Bessel beam 


\begin{equation}
J_0 (\rho k\,\, sin\theta ) \, {\rm exp} \Big\{i (zk \, 
cos\theta 
- \omega t ) \Big\}
\end{equation}
moving in the $z$ direction, where $\theta$ is a parameter around 20$^{\circ}$, 
$\rho$ is the cylindrical radial variable, $J_0$ is the Bessel function, 
and $\omega = kc$.

The authors found experimentally that signals moved in the $z$  
direction with speed equal to 

\begin{equation}
v = c / cos \theta \,,
\end{equation}
a superluminal speed that they assert agrees with the group velocity derived from
Eq.\ (1).

Putting aside the experimental result, I point out here that Eq.\ (2) is not
correct for the group velocity.  It  disagrees with the wave equation

\begin{equation}
{1\, \over c^2} \,\,{\partial ^2 \psi \over \partial t^2} = \nabla ^2 \psi \,,
\end {equation}
which gives a group velocity

\begin{equation}
v_g = c\,cos \theta
\end{equation}
for the Bessel beam of Eq.\ (1).  Therefore, the reported experimental result 
cannot be reconciled with the Maxwell equations.  If the experiment is correct, the Maxwell 
equations in the laboratory system are grossly in error at microwave frequencies,
independently of any considerations of relativity theory.

Consider a wave packet 

\begin{equation}
\psi (z,\rho ,t) = {1 \over \sqrt{2\pi }} \int k_{\rho} dk_{\rho} dk_z A(k_z ,k_{\rho} )
J_0 (\rho k_{\rho} ) \,{\rm exp} \Big\{i (zk_z - \omega (k_z , k_{\rho} )
t) \Big\} \,.
\end{equation}
(The two-dimensional transform is needed to represent a Bessel beam chopped into $z$ 
segments in spite of the fixed angle $\theta$ because $k_z$ is not constant.)  The wave
equation requires

\begin{equation}
\omega (k_z , k_{\rho} ) = c \sqrt {k{^2_z} + k {^2_{\rho}}} \,\,.
\end{equation}
The "center of gravity" of the wave packet is given by 

\begin{eqnarray}
\langle z \rangle _t={1\over 2\pi } \int k_{\rho} dk_{\rho} dk_z K_{\rho} dK_{\rho}
dK_z A (K_z , K_{\rho} ) ^* A(k_z , k_{\rho} ) {\rm exp}
\Big\{i\omega (K_z , K_{\rho} ) t \Big\} {\rm exp} \Big\{ -
i\omega (k_z , k_{\rho} ) t \Big\} \\
\nonumber  \times \,\Big[ \rho d\rho J_0 (K_{\rho} \rho ) J_0 (k_{\rho} \rho ) 
\Big] \times \Big[ dz\, {\rm exp} \{iK _z z \}z \,{\rm exp} \{ - ik_z z \} \Big ] \,.
\end{eqnarray}
Carrying out the $\rho$ and $z$ integrals gives ${1\over K_{\rho} }  
\delta (K_{\rho} - k_{\rho} ) (-2\pi i) \delta ' (K_z - k_z )$
and then carrying out
the two $K$ integrals gives

 \begin{equation}
\langle z \rangle _t = -i \int {\partial A(k_z , k_{\rho}) ^* \over \partial k_z }
A(k_z , k_{\rho} ) dk_z k_{\rho} dk_{\rho} + t \int {\partial \omega (k_z ,k_{\rho}) \over
\partial k_z}\, \Big|A(k_z , k_{\rho}) \Big| ^2 dk_z k_{\rho} dk_{\rho} = \langle z \rangle _0
+ \Big\langle v_g \Big\rangle t \,,
\end{equation}
where the group velocity is given by

\begin{equation}
v_g = {\partial \omega (k_z,k_{\rho}) \over \partial k_z} = {k_z c \over \sqrt{k_z^2 + k_{\rho}^2}} \,\leq c \,.
\end{equation}

For the Bessel beam of Ref.\ [1], the right-hand side of Eq.\ (9) is nearly equal 
to the constant $c\,cos\theta$  so a pulse of any shape moves with group velocity 
equal to $c\,cos\theta$ and not with the superluminal velocity given in Eq.\ (2).  In
particular, an initial delta-function pulse will move with the same group velocity.  In
the absence of dispersion, the delta function will not spread as it progresses, so the
signal velocity is also equal to $c\,cos\theta$.

It is perhaps useful to remark that superluminal propagation does not
necessarily follow from the data presented in Ref.\ [1] if the microwave
pulses do not have exactly the Bessel shape of the chopped Eq.\ (1).  A
chopped plane wave moving at angle $\theta$ to the $z$ axis will
activate detectors placed at intervals along that axis in such a way as
to give an illusion of superluminal propagation in the $z$ direction
with the velocity given in Eq.\ (2).  In the Bessel case, interference
between plane waves at all azimuthal angles removes the
``superluminal'' part of the wave.  However, incomplete cancellation in
 an imperfectly cylindrical wave could restore the illusion.

This work was supported by the U.S. Department of Energy, Nuclear Physics
 Division, under contract W-31-109-ENG-38.  I thank Mitio Inokuti, Pierre Lahaie, 
and David Y. Smith for helpful communications.

\section*{References}
[1] \,\,\,D. Mugnai, A. Ranfagni, and R. Ruggeri, Phys.\ Rev.\ Lett. $\bf 84$, 4830
(2000).

\end{document}